\documentclass[10pt]{article}
\usepackage{graphicx}
\usepackage{amsmath}
\usepackage{amssymb}
\usepackage{caption2}
\begin{document}
\bibliographystyle{prsty}
\begin{center}
{\large {\bf \sc{  The $B_c$ meson and its scalar cousin with  the QCD sum rules }}} \\[2mm]
Zhi-Gang Wang \footnote{E-mail,zgwang@aliyun.com.  }     \\
 Department of Physics, North China Electric Power University,
Baoding 071003, P. R. China
\end{center}

\begin{abstract}
In the present work, we use optical theorem to calculate the next-to-leading order corrections to  the QCD spectral densities directly in the QCD sum rules for the pseudoscalar and scalar $B_c$ mesons.  We take the experimental data as guides to perform updated analysis,  and obtain the masses and decay constants, especially the decay constants, which are the fundamental input parameters in the high energy physics,  therefore the pure leptonic decay widths, which can  be confronted to the experimental data in the future.
\end{abstract}



\section{Introduction}

In 1998, the CDF collaboration  observed the pseudoscalar  $B_c$ mesons  through  the semi-leptonic decay modes $B_c^{\pm} \to J/\psi \ell^{\pm}X $ and $B_c^{\pm} \to J/\psi \ell^{\pm}\bar{\nu}_{\ell} $ in the $p\bar{p}$ collisions at the energy $\sqrt{s}=1.8\,\rm{TeV}$ at the Fermilab Tevatron, the measured mass is $6.40 \pm 0.39 \pm 0.13 \,\rm{GeV}$ \cite{CDF1998-1,CDF1998-2}. It is the first time to observe the bottom-charm meson experimentally.

In 2007, the  CDF collaboration  confirmed  the  $B_c$ mesons through the non-leptonic decay modes $B_c^{\pm} \to J/\psi \pi^{\pm}$  with the   measured mass $6275.6 \pm 2.9\pm 2.5 \, \rm{MeV}$ \cite{CDF2008}.
In 2008, the D0 collaboration reconstructed the non-leptonic decays $B_c^{\pm} \to J/\psi \pi^{\pm}$  and confirmed  the $B_c$ mesons  with the measured  mass  $6300 \pm 14 \pm 5\,
\rm{ MeV}$ \cite{D02008}. Now the $B_c$ meson is well established, the average value listed in the  Review of Particle Physics is  $6274.47\pm 0.27\pm 0.17\,\rm{MeV}$  \cite{PDG}.

In 2014, the ATLAS collaboration reported the observation of a structure in the $B_c^\pm \pi^+\pi^-$ invariant mass spectrum  with a significance of 5.2 standard deviations, which is consistent with the predicted $B_c^\prime$ meson with a mass of $6842\pm4\pm 5\, \rm{MeV}$ \cite{ATLAS-2014}.

In 2019, the CMS collaboration observed two excited $\bar{b}c$ states  in the $B^+_c\pi^+\pi^-$ invariant mass spectrum with a significance exceeding five standard deviations, which are  consistent with the $B^{\prime+}_c$ and $B_c^{*\prime+}$,  respectively \cite{CMS-2019}. The two states are separated in mass by $29.1 \pm 1.5 \pm 0.7 \, \rm{MeV}$, and the mass of the $B^{\prime+}_c$  is measured to be $6871.0 \pm 1.2 \pm 0.8 \pm 0.8 \,\rm{ MeV}$.
Also in 2019, the LHCb collaboration observed the excited $B^{\prime+}_c$ (with a
global (local) statistical significance of $2.2\sigma$ ($3.2\sigma$)) and $B_c^{*\prime+}$ (with a global (local) statistical significance of
$6.3\sigma$ ($6.8\sigma$)) mesons in the $B^+_c\pi^+\pi^-$ invariant mass spectrum.
  The $B_c^{*\prime+}$ meson has a mass of
  $6841.2 \pm 0.6  \pm 0.1  \pm 0.8 \,\rm{MeV}$, which is reconstructed
  without the low-energy photon emitted in the $B_c^{*+} \to B_c^+ \gamma$ decay
  following the process $B_c^{\prime*+} \to B_c^{*+} \pi^+ \pi^-$, while the $B^{\prime+}_c$ meson has   a mass of
  $6872.1 \pm 1.3  \pm 0.1  \pm 0.8 \,\rm{MeV}$ \cite{LHCb-2019}.

It is very odd that the $B_c^\prime$ meson emerges as heavier than the mass of the  $B_c ^{*\prime}$ meson, which is in conflict with all the theoretical estimations, this maybe or maybe not due to impossibility of reconstruction of the low-energy photon in the $ B_c ^{\ast +}\rightarrow B_c ^{+} \gamma $ decay \cite{WZG-Bc-gamma}, more precisely experimental data are still needed.
Only the $B_c$ and $B_c^\prime$ mesons are listed in Review of Particle Physics \cite{PDG}, which are
in contrary to the copious (well-established) spectroscopy  of the  charmonium  and bottomonium states. Despite the enormous  developments on the heavy quark physics in recent years, the bottom-charm spectroscopy remains poorly known, which calls for further investigations.

The  beauty-charm  mesons provide an optimal laboratory for exploring both the perturbative and nonperturbative dynamics of the heavy quarks, due to absence  of the light quark's  contamination, and  for exploring the strong and electro-weak interactions,
as they are composed of two different heavy flavor quarks and cannot annihilate into gluons or photons.
For the  excited $c \bar b$ states, which lie below the $BD$ threshold,
would decay into the  $B_c$ meson through the
radiative decays or hadronic decays \cite{GI-1,GI-2}, while the ground state $B_c$  can only  decay weakly through emitting a virtual $W$-boson, thus it cannot decay through  strong or electromagnetic  interactions.

There have been a number of theoretical works on the mass spectroscopy of the bottom-charm  mesons, such as
   the relativized (or relativistic) quark model with an special  potential \cite{GI-1,GI-2,RPMD-EFG,RPMD-GJ,RPMD-ZVR}, the  nonrelativistic quark model with  an special potential \cite{NRPMD-SR-GKLT-1,NRPMD-SR-GKLT-2,NRPMD-EQ-1,NRPMD-EQ,NRPMD-Monteiro,NRPMD-Fulc,NRPMD-Soni,NRPMD-Ortega,NRPMD-LuQF},  the semi-relativistic quark model using the shifted large-$N$ expansion \cite{LargeN-IS2004,Ikhdair2006}, the perturbative QCD \cite{pQCD-BV2000}, the nonrelativistic renormalization group \cite{RGroup-Penin2004}, the lattice QCD \cite{Latt0909-Gregory-PRL,Latt-Davies,Latt-Mathur-mass,Jones-latt}, the Bethe-Salpeter equation \cite{BSE-Vary,BSE-WangGL-1,BSE-WangGL-2},
 the full QCD sum rules \cite{QCDSR-Bagan1994,QCDSR-Chabab1994,QCDSR-Colangelo1993,QCDSR-Kiselev1993,WZG-APPB-Bc,QCDSR-WZG-BcV,
 QCDSR-Aliev-Bc}, the potential model  combined with the QCD sum rules \cite{NRPMD-SR-GKLT-1,NRPMD-SR-GKLT-2}, etc.

With the continuous developments  in experimental techniques, we  expect that more $c\bar b$ states  would  be observed by the ATLAS, CMS, LHCb, etc in the future.
The decay constant, which parameterizes the coupling between a current and a meson, plays an important role in exploring  the exclusive processes, because the decay constants  are not only a fundamental parameter describing the pure leptonic
decays, but also are an universal input parameter related to the distribution amplitudes, form-factors, partial decay widths and branching fractions  in many  processes. By precisely measuring the branching fractions, we can resort to the decay constants to extract the CKM matrix element in the standard model  and search for new physics beyond the standard model \cite{NewP-Blanke}.

Decay constants of the bottom-charm mesons have been investigated  in a number of  theoretical  approaches, such as  the full QCD sum rules \cite{QCDSR-Bagan1994,QCDSR-Chabab1994,QCDSR-Colangelo1993,QCDSR-Kiselev1993,WZG-APPB-Bc,
QCDSR-WZG-BcV,QCDSR-Aliev-Bc,QCDSR-Narison-Bc,QCDSR-Schilcher-Bc}, the potential model combined with the QCD sum rules \cite{NRPMD-SR-GKLT-1,NRPMD-SR-GKLT-2}, the QCD sum rule combined with the heavy quark effective theory \cite{Decay-SR-HQEFT-Onishchenko,Decay-SR-HQEFT-Lee,Decay-SR-HQEFT-XiaoZJ,Decay-SR-HQEFT-ZhuRL,
Decay-SR-HQEFT-ZhuRL-2,Decay-SR-HQEFT-QiaoCF,Decay-SR-HQEFT-Sang,Decay-SR-HQEFT-Feng},  the covariant light-front quark model \cite{Decay-LFQM-Verma,Decay-LFQM-Tang},  the lattice non-relativistic QCD \cite{Jones-latt},
     the  shifted large-$N$ expansion method \cite{Ikhdair2006},   the field correlator method \cite{BBS}, etc. However, the values from different theoretical approaches vary in a large range, it is interesting and necessary to extend our previous works on the vector and axialvector $B_c$ mesons  \cite{QCDSR-WZG-BcV} to   investigate  the pseudoscalar and scalar $B_c$ mesons with the full QCD sum rules by including next-to-leading order radiative corrections and choose the  updated input parameters, thus our investigations are performed in a consistent and systematic way. We take the experimental data \cite{PDG,ATLAS-2014,CMS-2019,LHCb-2019} as guides to choose the suitable Borel parameters and continuum threshold parameters,  examine  the masses and decay constants of the pseudoscalar  and scalar $B_c$ mesons with the full QCD sum rules, therefore  we calculate the pure leptonic decay widths to be confronted to the experimental data in the future.

The article is arranged:  we calculate the next-to-leading order contributions to the spectral densities and obtain the QCD sum rules in Sect.2;
in Sect.3, we present the numerical results and discussions; and Sect.4 is reserved for our conclusions.

\section{Explicit calculations of the QCD spectral densities at the next-to-leading order}
We write down  the two-point correlation functions firstly,
\begin{eqnarray}
\Pi_{P/S}(p^2)&=&i\int d^4x e^{ip \cdot x} \langle 0|T\left\{J(x)J^{\dagger}(0)\right\}|0\rangle \, ,
\end{eqnarray}
where  $J(x)=J_P(x)$ and $J_S(x)$,
\begin{eqnarray}
J_P(x)&=&\bar{c}(x)i\gamma_5 b(x)\, , \nonumber\\
J_S(x)&=&\bar{c}(x) b(x)\, ,
\end{eqnarray}
the subscripts $P$ and $S$ represent the pseudoscalar and scalar mesons, respectively.
 The correlation functions can be written in the form,
 \begin{eqnarray}
\Pi_{P/S}(p^2)&=&\frac{1}{\pi}\int_{(m_b+m_c)^2}^\infty ds \frac{{\rm Im\Pi}_{P/S}(s)}{s-p^2}\, ,
\end{eqnarray}
  according to the dispersion relation, where
\begin{eqnarray}
\frac{{\rm Im}\Pi_{P/S}(s)}{\pi}&=&\rho_{P/S}(s) \nonumber\\
&=&\rho^0_{P/S}(s)+\rho^1_{P/S}(s)+\rho^2_{P/S}(s)+\cdots\, ,
\end{eqnarray}
the QCD spectral densities $\rho_{P/S}(s)$ are expanded in terms of the strong fine structure constant $\alpha_s=\frac{g_s^2}{4\pi}$, the $\rho^0_{P/S}(s)$, $\rho^1_{P/S}(s)$, $\rho^2_{P/S}(s)$, $\cdots$ are the spectral densities of the leading order, next-to-leading order, and next-to-next-to-leading order, $\cdots$. At the leading order,
\begin{eqnarray}
  \rho^0_{P/S}(s)&=& \frac{3}{8\pi^2}\frac{\sqrt{\lambda(s,m_b^2,m_c^2)}}{s}\left[s-(m_b\mp m_c)^2 \right]\,, \end{eqnarray}
where the standard phase space factor,
\begin{eqnarray}
\lambda(s,m_b^2,m_c^2)&=&s^2+m_b^4+m_c^4-2sm_b^2-2sm_c^2-2m_b^2m_c^2\, .
\end{eqnarray}

\begin{figure}
 \centering
 \includegraphics[totalheight=2.7cm,width=14cm]{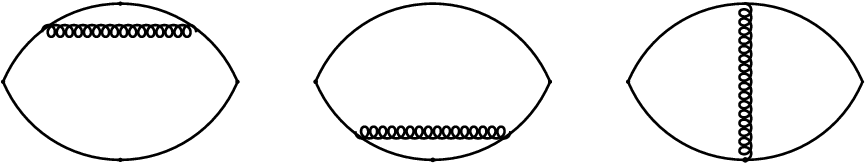}
    \caption{The next-to-leading order contributions to the correlation functions. }\label{two-loop}
\end{figure}

At the next-to-leading order, there exist three standard Feynman diagrams, which correspond to the self-energy and vertex corrections respectively,  and make contributions  to the correlation functions, see Fig.\ref{two-loop}.
We calculate the imaginary parts of those Feynman diagrams  resorting to  the Cutkosky's rule or optical theorem, the two methods  result in the same analytical expressions,  then we use dispersion relation to acquire the correlation functions at the quark-gluon level \cite{WZG-APPB-Bc,Reinders85}. There exist  ten
possible cuts, six cuts  make attributions to virtual gluon emissions  and four cuts make attributions  to real gluon emissions.

\begin{figure}
 \centering
 \includegraphics[totalheight=3.7cm,width=14cm]{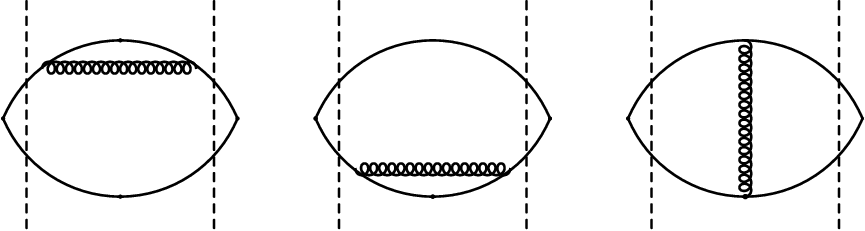}
    \caption{Six possible cuts correspond to virtual gluon emissions. }\label{two-virt}
\end{figure}

\begin{figure}
 \centering
 \includegraphics[totalheight=2cm,width=8cm]{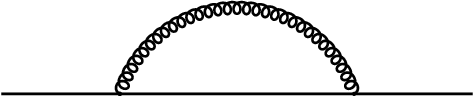}
    \caption{The quark self-energy correction. }\label{Self-Energy}
\end{figure}

\begin{figure}
 \centering
 \includegraphics[totalheight=3cm,width=4cm]{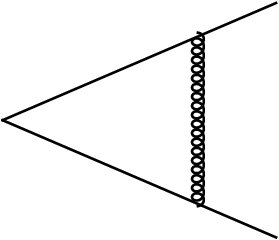}
    \caption{The vertex correction. } \label{Vertex}
\end{figure}

The six cuts which are shown in Fig.\ref{two-virt} make attributions  to virtual gluon emissions, and could be  classified as  the self-energy and vertex corrections, respectively.
We calculate the Feynman diagrams straightforwardly   by adopting the dimensional regularization to regularize both the ultraviolet and infrared divergences,  and
resort to the on-shell renormalization scheme to absorb  the ultraviolet divergences by accomplishing  the
wave-function and quark-mass renormalizations.
Then we take account of all  contributions which are shown in Fig.\ref{two-virt} by the simple replacements of the vertexes in all the currents,
\begin{eqnarray}
\bar{u}(p_1)i\gamma_5 u(p_2)&\rightarrow& \bar{u}(p_1)i\gamma_5 u(p_2)+\bar{u}(p_1)i\widetilde{\Gamma}_5 u(p_2)   \nonumber \\
&=& \sqrt{Z_1}\sqrt{Z_2}\bar{u}(p_1)i\gamma_5 u(p_2)+\bar{u}(p_1)i\Gamma_5 u(p_2)\nonumber \\
&=& \bar{u}(p_1)i\gamma_5 u(p_2)\left(1+\frac{1}{2}\delta Z_1+\frac{1}{2}\delta Z_2 \right)+\bar{u}(p_1)i\Gamma_5 u(p_2)\, ,
\end{eqnarray}
\begin{eqnarray}
\bar{u}(p_1) u(p_2)&\rightarrow& \bar{u}(p_1) u(p_2)+\bar{u}(p_1)\widetilde{\Gamma}_0 u(p_2)   \nonumber \\
&=& \sqrt{Z_1}\sqrt{Z_2}\bar{u}(p_1) u(p_2)+\bar{u}(p_1)\Gamma_0 u(p_2)\nonumber \\
&=& \bar{u}(p_1) u(p_2)\left(1+\frac{1}{2}\delta Z_1+\frac{1}{2}\delta Z_2 \right)+\bar{u}(p_1)\Gamma_0 u(p_2)\, ,
\end{eqnarray}
where
\begin{eqnarray}
Z_i&=&1+\delta Z_{i}=1+\frac{4}{3}\frac{\alpha_s}{\pi}\left(-\frac{1}{4\varepsilon_{\rm UV}}+\frac{1}{2\varepsilon_{\rm IR}}+\frac{3}{4}\log\frac{m_i^2}{4\pi\mu^2}+\frac{3}{4}\gamma-1\right)\, ,
\end{eqnarray}
is the $i$ quark's wave-function renormalization constant which originates from the self-energy diagram, see Fig.\ref{Self-Energy},
and
\begin{eqnarray}
\Gamma_{5/0} &=&\gamma_5 \frac{4}{3}g_s^2\int_0^1 dx \int_0^{1-x}dy \int \frac{d^D k_E}{(2\pi)^D} \nonumber\\
&& \frac{\Gamma(3)}{\left[k^2_E+(x p_1+y p_2)^2\right]^3}\Big\{4k^2_E\Big(1-\frac{1}{2}\varepsilon_{\rm UV} \Big)+2(1-x-y+2xy)(s-m_b^2-m_c^2) \nonumber\\
&&\pm2(x+y)m_bm_c+2x(1-2x)m_b^2+2y(1-2y)m_c^2\Big\} \, ,
\end{eqnarray}
for the vertex diagrams after accomplishing the Wick's rotation, see Fig.\ref{Vertex}, where the $\gamma$ is the Euler constant, the $\mu$ is the energy scale of renormalization, and the $k_E=(k_1,k_2,k_3,k_4)$ is Euclidean four-momentum.  We set the dimension  $D=4-2\varepsilon_{\rm UV}=4+2\varepsilon_{\rm IR}$ to regularize the ultraviolet and infrared divergences respectively, where the $\varepsilon_{\rm UV}$ and $\varepsilon_{\rm IR}$ are  positive dimension-less quantities,  and we would add the  energy scale factors $\mu^{ 2\varepsilon_{\rm UV}}$ or $\mu^{ -2\varepsilon_{\rm IR}}$ if necessary.

We accomplish  all the integrals over all the variables, and observe that the  ultraviolet divergences $\frac{1}{\varepsilon_{\rm UV}}$ in the $\Gamma_{5/0}$, $\delta Z_1$ and $\delta Z_2$ are canceled out completely with each other, the offsets  are warranted by  the Ward identity. So the total contributions do not have ultraviolet divergences,
\begin{eqnarray}
\widetilde{\Gamma}_5 &=& \frac{4}{3}\frac{\alpha_s}{4\pi}\gamma_5 f_P(s) \, , \nonumber \\
\widetilde{\Gamma}_0 &=& \frac{4}{3}\frac{\alpha_s}{4\pi}  f_S(s) \, ,
\end{eqnarray}
where
\begin{eqnarray}
f_{P/S}(s) &=& \overline{f}_{P/S}(s)+ \frac{2}{\varepsilon_{\rm IR}} +3\log\frac{m_bm_c}{4\pi\mu^2} +4\log\frac{4\pi\mu^2}{s}-\gamma+4-\frac{2(s-m_b^2-m_c^2)}{\sqrt{\lambda(s,m_b^2,m_c^2)}}\log\left(\frac{1+\omega}{1-\omega}\right)  \nonumber\\
&& \left(\frac{1}{\varepsilon_{\rm IR}} +\log\frac{s}{4\pi\mu^2}+\gamma\right)\, ,\nonumber\\
\overline{f}_{P/S}(s)&=& 4\overline{V}(s)+2(s-m_b^2-m_c^2)\left[\overline{V}_{00}(s)-V_{10}(s)-V_{01}(s)+2V_{11}(s)\right] \pm 2m_bm_c\nonumber\\
&&\left[ V_{10}(s)+V_{01}(s)\right]+2m_b^2\left[ V_{10}(s)-2V_{20}(s)\right]+2m_c^2\left[ V_{01}(s)-2V_{02}(s)\right]\, ,\nonumber\\
\omega&=&\sqrt{\frac{s-(m_b+m_c)^2}{s-(m_b-m_c)^2}}\, ,
\end{eqnarray}
and $s=p^2$, the definitions and explicit expressions of the notations $\overline{V}(s)$, $\overline{V}_{00}(s)$ and $V_{ij}(s)$ with $i,j=0,1,2$ are given  in the appendix.

The contributions of all the  virtual gluon emissions to the imaginary parts of the Feynman diagrams in Fig.\ref{two-loop} are,
\begin{eqnarray}
\frac{{\rm Im}\Pi^V_{P/S}(s)}{\pi}&=&\frac{4}{3}\frac{\alpha_s}{4\pi}\frac{6}{\pi}\int \frac{d^{D-1}\vec{p}_1}{(2\pi)^{D-1}2E_{p_1}}\frac{d^{D-1}\vec{p}_2}{(2\pi)^{D-1}2E_{p_2}}(2\pi)^D \delta^D(p-p_1-p_2) f(s)\left[s-(m_b\mp m_c)^2\right]\, , \nonumber\\
\end{eqnarray}
the superscript $V$ denotes the virtual gluon emissions.
We accomplish all the integrals  straightforwardly  in the dimension $D=4+2\varepsilon_{\rm IR}$  as there does not  exist  ultraviolet divergence, and obtain the analytical expressions,
\begin{eqnarray}
\frac{{\rm Im}\Pi^V_{P/S}(s)}{\pi}&=&\frac{4}{3}\frac{\alpha_s}{\pi}\rho_{P/S}^0(s)\left\{\frac{1}{\varepsilon_{\rm IR}}-2\log4\pi+\frac{1}{2}\gamma+\frac{1}{2}\log\frac{\lambda^2(s,m_b^2,m_c^2)m_b^3m_c^3}{\mu^8 s^3} +\frac{1}{2}\overline{f}_{P/S}(s)\right.\nonumber\\
&&\left.-\frac{s-m_b^2-m_c^2}{\sqrt{\lambda(s,m_b^2,m_c^2)}}\log\left(\frac{1+\omega}{1-\omega}\right)\left[\frac{1}{\varepsilon_{\rm IR}}-2\log4\pi+2\gamma-2+\log\frac{\lambda(s,m_b^2,m_c^2)}{\mu^4} \right] \right\}  \, .\nonumber\\
\end{eqnarray}

\begin{figure}
 \centering
 \includegraphics[totalheight=3.7cm,width=14cm]{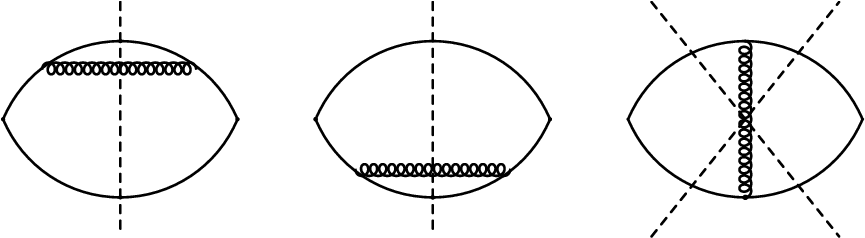}
    \caption{Four possible cuts correspond to  real gluon emissions. }\label{two-real}
\end{figure}

\begin{figure}
 \centering
 \includegraphics[totalheight=3cm,width=7cm]{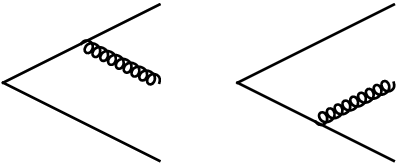}
    \caption{The amplitudes for the real gluon emissions. }\label{Real-Gluons}
\end{figure}

The four cuts in the Feynman diagrams shown in Fig.\ref{two-real} only make contributions to the real gluon emissions, the  corresponding  scattering amplitudes  are shown explicitly in Fig.\ref{Real-Gluons}. From the two diagrams in Fig.\ref{Real-Gluons},
we  write down the scattering amplitudes $T^a_{5,\alpha}(p)$ and $T^a_{0,\alpha}(p)$ ,
\begin{eqnarray}
T^a_{5,\alpha}(p)&=&\bar{u}(p_1)\left\{ ig_s \frac{\lambda^a}{2}\gamma_\alpha \frac{i}{\!\not\!{p}_1+\!\not\!{k}-m_b}i\gamma_5+i\gamma_5\frac{i}{-\!\not\!{p}_2-\!\not\!{k}-m_c}ig_s\frac{\lambda^a}{2}\gamma_\alpha\right\}v(p_2)\,  , \nonumber\\
T^a_{0,\alpha}(p)&=&\bar{u}(p_1)\left\{ ig_s \frac{\lambda^a}{2}\gamma_\alpha \frac{i}{\!\not\!{p}_1+\!\not\!{k}-m_b}+\frac{i}{-\!\not\!{p}_2-\!\not\!{k}-m_c}ig_s\frac{\lambda^a}{2}\gamma_\alpha\right\}v(p_2)\,  ,
\end{eqnarray}
where the $\lambda^a$ is the Gell-Mann matrix. Then we obtain the contributions  to the imaginary parts  of the Feynman diagrams  with the optical theorem,
\begin{eqnarray}
\frac{{\rm Im}\Pi^R_{P/S}(s)}{\pi}&=&-\frac{1}{2\pi}  \int \frac{d^{D-1}\vec{k}}{(2\pi)^{D-1}2E_k}\frac{d^{D-1}\vec{p}_1}{(2\pi)^{D-1}2E_{p_1}}\frac{d^{D-1}\vec{p}_2}{(2\pi)^{D-1}2E_{p_2}}(2\pi)^{D}\delta^D(p-k-p_1-p_2)\nonumber\\
&&{\rm Tr}\left\{T^a_{5/0,\alpha}(p)T^{a\dagger}_{5/0,\beta}(p)\right\}g^{\alpha\beta} \nonumber\\
&=&-\frac{2g_s^2}{\pi}  \int \frac{d^{D-1}\vec{k}}{(2\pi)^{D-1}2E_k}\frac{d^{D-1}\vec{p}_1}{(2\pi)^{D-1}2E_{p_1}}\frac{d^{D-1}\vec{p}_2}{(2\pi)^{D-1}2E_{p_2}}(2\pi)^{D}\delta^D(p-k-p_1-p_2)\nonumber\\
&&\left\{2\left[ s-(m_b\mp m_c)^2\right]\left[\frac{m_b^2}{(k\cdot p_1)^2} +\frac{m_c^2}{(k\cdot p_2)^2}-\frac{s-m_b^2-m_c^2}{k\cdot p_1 k\cdot p_2}\right.\right.\nonumber\\
&&\left.\left.+\frac{s-K^2}{k\cdot p_1 k\cdot p_2}\right]  -\frac{(s-K^2)^2}{k\cdot p_1 k\cdot p_2}\right\} \, ,
\end{eqnarray}
where we have used the formulas $\sum u(p_1)\bar{u}(p_1)=\!\not\!{p}_1+m_b$ and $\sum v(p_2)\bar{v}(p_2)=\!\not\!{p}_2-m_c$ for the quark and antiquark respectively, and we introduce the symbol $K^2=(p_1+p_2)^2$ for simplicity, and introduce the superscript $R$ to denote the real gluon emissions. We accomplish  the integrals  in the dimension $D=4+2\varepsilon_{\rm IR}$ because there only exist the infrared divergences (no ultraviolet divergences), and obtain the contributions,
\begin{eqnarray}
\frac{{\rm Im}\Pi^R_{P/S}(s)}{\pi}&=&\frac{4}{3}\frac{\alpha_s}{\pi}\rho_{P/S}^0(s)\left\{-\frac{1}{\varepsilon_{\rm IR}}+2\log4\pi-2\gamma+2-\log\frac{\lambda^3(s,m_b^2,m_c^2)}{m_b^2m_c^2s^2\mu^4} +(s-m_b^2-m_c^2)\overline{R}_{12}(s)\right.\nonumber\\
&&-\overline{R}_{11}(s)-\overline{R}_{22}(s)-R_{12}^1(s) +\frac{R_{12}^2}{2}\frac{1}{s-(m_b\mp m_c)^2}+\frac{s-m_b^2-m_c^2}{\sqrt{\lambda(s,m_b^2,m_c^2)}} \nonumber\\
&&\left.\log\left(\frac{1+\omega}{1-\omega}\right) \left[\frac{1}{\varepsilon_{\rm IR}}-2\log4\pi+2\gamma-2+\log\frac{\lambda^3(s,m_b^2,m_c^2)}{m_b^2m_c^2s^2\mu^4} \right]\right\}\, ,
\end{eqnarray}
the definitions and  explicit expressions of the $\overline{R}_{11}(s)$, $\overline{R}_{22}(s)$, $\overline{R}_{12}(s)$, $R_{12}^1(s)$ and $R^2_{12}(s)$ are  given  in the appendix.

Now we  obtain the total QCD spectral densities at the next-to-leading order,
\begin{eqnarray}
\rho_{P/S}^1(s)&=&\frac{4}{3}\frac{\alpha_s}{\pi}\rho_{P/S}^0(s)\left\{\frac{1}{2}\overline{f}_{P/S}(s)
-\overline{R}_{11}(s)-\overline{R}_{22}(s)-R_{12}^1(s)+(s-m_b^2-m_c^2)\,\overline{R}_{12}(s)\right.\nonumber\\
&&\left.+\frac{R_{12}^2}{2}\frac{1}{s-(m_b\mp m_c)^2}
-\frac{3}{2}\gamma+2 +\frac{1}{2}\log\frac{m_b^7m_c^7s}{\lambda^4(s,m_b^2,m_c^2)} \right.\nonumber\\
&&\left.+\frac{s-m_b^2-m_c^2}{\sqrt{\lambda(s,m_b^2,m_c^2)}}\log\left(\frac{1+\omega}{1-\omega}\right)
\log\frac{\lambda^2(s,m_b^2,m_c^2)}{m_b^2m_c^2s^2} \right\}  \, .
\end{eqnarray}
The infrared divergences of the forms $\frac{1}{\varepsilon_{\rm IR}}$, $\log\left(\frac{1+\omega}{1-\omega} \right)\frac{1}{\varepsilon_{\rm IR}}$   from the virtual and real gluon emissions  are canceled out with each other completely, the offsets are guaranteed by the Lee-Nauenberg theorem \cite{LeeBook}. The analytical expressions are applicable in  many phenomenological analysis besides  the QCD sum rules.

Then we calculate the contributions of the gluon condensate directly, the calculations are easy and no much to say. Finally, we obtain the analytical expressions of the QCD spectral densities,  take the
quark-hadron duality below the continuum thresholds $s^0_{P/S}$ and perform the Borel transforms  in regard  to the variable
$P^2=-p^2$ to acquire  the QCD sum rules,
\begin{eqnarray}\label{QCDSR}
\frac{f_{P/S}^2 M_{P/S}^4}{(m_b\pm m_c)^2} \exp\left(-\frac{M_{P/S}^2}{T^2}\right)&=& \int_{(m_b+m_c)^2}^{s^0_{P/S}} ds\left[\rho^0_{P/S}(s)+\rho^{1}_{P/S}(s)+\rho^{\rm con}_{P/S}(s)\right] \exp\left(-\frac{s}{T^2}\right) \, ,\nonumber \\
\end{eqnarray}
where
\begin{eqnarray}
\rho^{\rm con}_{P/S}(s)&=&\mp \frac{m_b m_c}{24T^4}\langle\frac{\alpha_sGG}{\pi}\rangle\int_0^1dx
\left[ \frac{m_c^2}{x^3}+\frac{m_b^2}{(1-x)^3}\right]\delta(s-\widetilde{m}_Q^2) \nonumber\\
&&\pm \frac{m_b m_c}{8T^2}\langle\frac{\alpha_sGG}{\pi}\rangle\int_0^1dx\left[ \frac{1}{x^2}+\frac{1}{(1-x)^2}\right]  \delta(s-\widetilde{m}_Q^2)\nonumber\\
&&-\frac{s}{24T^4}\langle\frac{\alpha_sGG}{\pi}\rangle\int_0^1dx
\left[ \frac{(1-x)m_c^2}{x^2}+\frac{x m_b^2}{(1-x)^2}\right]\delta(s-\widetilde{m}_Q^2) \, ,
\end{eqnarray}
 $\widetilde{m}_Q^2=\frac{m_b^2}{1-x}+\frac{m_c^2}{x}$, the $T^2$ is the Borel parameter, and the
 decay constants are defined by,
 \begin{eqnarray}\label{decay-const-defin-1}
 \langle 0|J_P(0)|P(p)\rangle&=&\frac{f_PM_P^2}{m_b+m_c}\, , \nonumber \\
 \langle 0|J_S(0)|S(p)\rangle&=&\frac{f_SM_S^2}{m_b-m_c}\, ,
 \end{eqnarray}
in other words,
\begin{eqnarray}
 \langle 0|J^\alpha_A(0)|P(p)\rangle&=&if_P\,p^\alpha\, , \nonumber \\
 \langle 0|J_V^\alpha(0)|S(p)\rangle&=&if_S\,p^\alpha\, ,
\end{eqnarray}
the subscripts $A$ and $V$ denote the axial-vector and vector currents, respectively.

 We eliminate the decay constants $f_{P/S}$ and obtain the QCD sum rules for the masses of the  pseudoscalar and scalar $B_c$ mesons,
 \begin{eqnarray}
 M_{P/S}^2&=& \frac{\int_{(m_b+m_c)^2}^{s^0_{P/S}} ds\frac{d}{d \left(-1/T^2\right)}\left[\rho^0_{P/S}(s)+\rho^{1}_{P/S}(s)+\rho^{\rm con}_{P/S}(s)\right]\exp\left(-\frac{s}{T^2}\right)}{\int_{(m_b+m_c)^2}^{s^0_{P/S}} ds \left[\rho^0_{P/S}(s)+\rho^{1}_{P/S}(s)+\rho^{\rm con}_{P/S}(s)\right]\exp\left(-\frac{s}{T^2}\right)}\, .
\end{eqnarray}

\section{Numerical results and discussions}
 The value of the gluon condensate $\langle \frac{\alpha_s
GG}{\pi}\rangle $ has been updated from time to time, and changes
greatly, we adopt  the updated value $\langle \frac{\alpha_s GG}{\pi}\rangle=0.022 \pm
0.004\,\rm{GeV}^4 $ \cite{Narison-gc-1105}.
We take the $\overline{MS}$ masses of the heavy  quarks
$m_{c}(m_c)=1.275\pm0.025\,\rm{GeV}$ and $m_{b}(m_b)=4.18\pm0.03\,\rm{GeV}$
 from the Particle Data Group \cite{PDG}.
In addition,  we take  account of the energy-scale dependence of the $\overline{MS}$ masses,
 \begin{eqnarray}
 m_Q(\mu)&=&m_Q(m_Q)\left[\frac{\alpha_{s}(\mu)}{\alpha_{s}(m_Q)}\right]^{\frac{12}{33-2n_f}} \, ,\nonumber\\
\alpha_s(\mu)&=&\frac{1}{b_0t}\left[1-\frac{b_1}{b_0^2}\frac{\log t}{t} +\frac{b_1^2(\log^2{t}-\log{t}-1)+b_0b_2}{b_0^4t^2}\right]\, ,
\end{eqnarray}
  where $t=\log \frac{\mu^2}{\Lambda^2}$, $b_0=\frac{33-2n_f}{12\pi}$, $b_1=\frac{153-19n_f}{24\pi^2}$, $b_2=\frac{2857-\frac{5033}{9}n_f+\frac{325}{27}n_f^2}{128\pi^3}$,  $\Lambda=213\,\rm{MeV}$, $296\,\rm{MeV}$  and  $339\,\rm{MeV}$ for the quark flavor numbers  $n_f=5$, $4$ and $3$, respectively  \cite{PDG}.
We choose $n_f=4$ and $5$  for the $c$ and $b$ quarks, respectively, and then
evolve all the heavy quark masses  to the  typical energy scale $\mu=2\,\rm{GeV}$.

The lower threshold $(m_b+m_c)^2$  in the QCD sum rules in Eq.\eqref{QCDSR} decreases quickly with increase of the energy scale,  the energy scale should be larger than $1.7\,\rm{GeV}$, which corresponds to the  squared mass of the $B_c$ meson, $39.4\,\rm{GeV}^2$. If we take the typical energy scale  $\mu=2\,\rm{GeV}$, which corresponds to the lower threshold $(m_b+m_c)^2\approx 36.0\,{\rm{GeV}}^2< M_P^2$, it is reasonable and feasible to choose such a particular energy scale.

The experimental  masses of the  $B_c$ and $B_c^\prime$ mesons are  $6274.47\pm 0.27\pm 0.17\,\rm{MeV}$ and $6871.2\pm 1.0\,\rm{MeV}$ respectively from the Particle Data Group \cite{PDG}. The scalar $B_c$ meson  still escapes the experimental detection, roughly speaking, the theoretical mass  is $6712\pm 18\pm 7\,\rm{MeV}$ from the lattice QCD \cite{Latt-Mathur-mass} or $6714\,\rm{MeV}$ from the nonrelativistic quark model \cite{NRPMD-LuQF}.
We can tentatively take
  the continuum threshold parameters as $s^0_{P}=(39-47)\,\rm{GeV}^2$ and $s^0_{S}=(45-55)\,\rm{GeV}^2$, and search for the ideal values by assuming the energy gap between the ground state and first radial excited states is about $0.6\,\rm{GeV}$, if lacking experimental data; we always resort to such an  assumption in the QCD sum rules.

\begin{figure}
 \centering
 \includegraphics[totalheight=5cm,width=6cm]{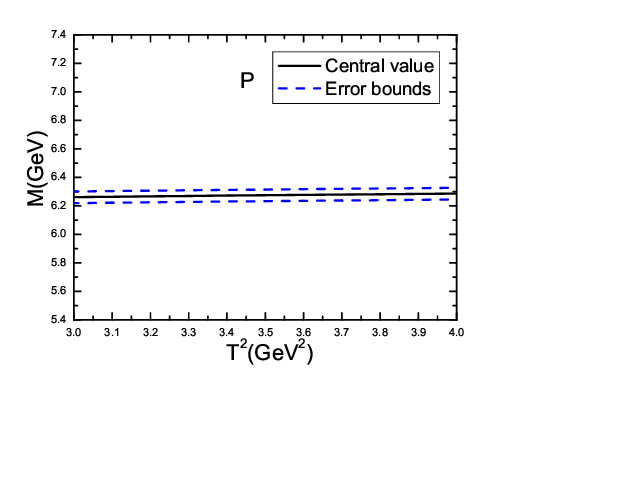}
\includegraphics[totalheight=5cm,width=6cm]{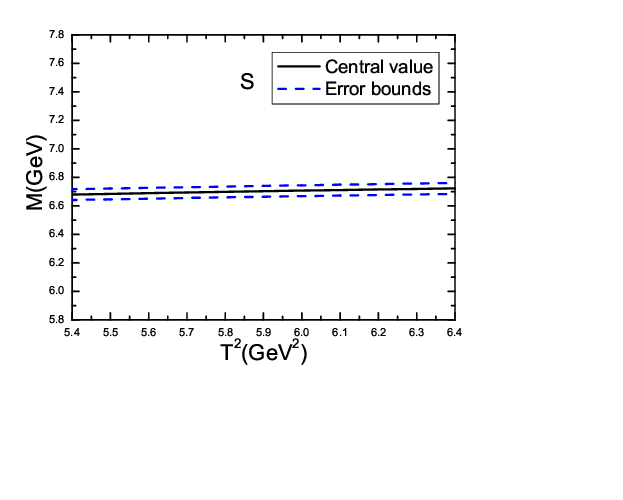}
        \caption{ The masses  of the pseudoscalar ($P$) and scalar ($S$) $B_c$ mesons with variations of the Borel parameters $T^2$. } \label{mass-Borel}
\end{figure}

\begin{figure}
 \centering
 \includegraphics[totalheight=5cm,width=6cm]{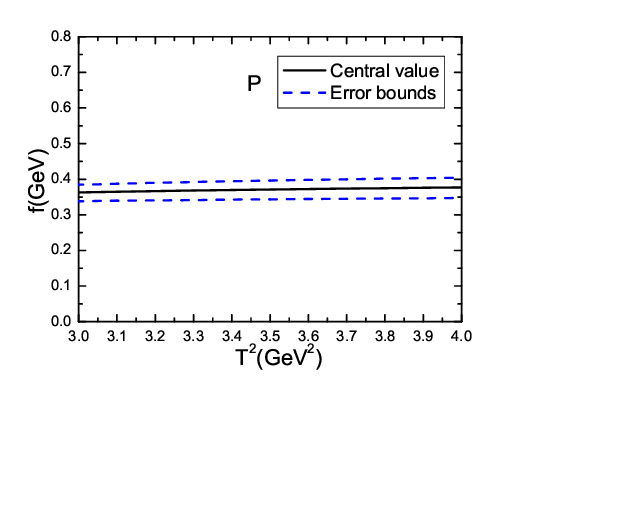}
\includegraphics[totalheight=5cm,width=6cm]{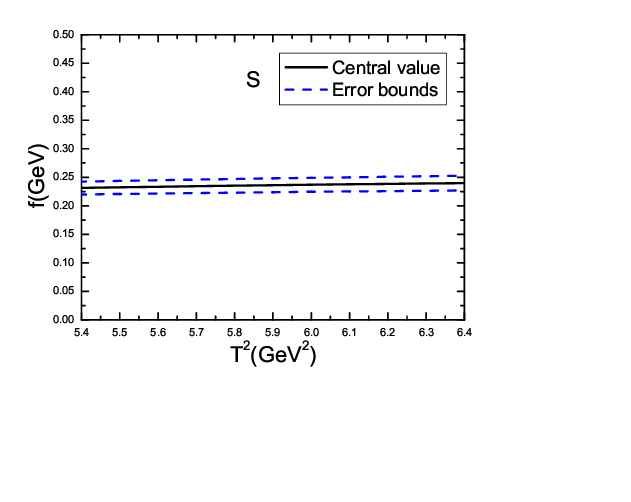}
        \caption{ The decay constants  of the pseudoscalar ($P$) and scalar ($S$) $B_c$ mesons with variations of the Borel parameters $T^2$. } \label{residue-Borel}
\end{figure}

After trial and error, we obtain the ideal Borel windows and continuum threshold parameters,  and
 the corresponding pole contributions about $(70-85)\%$, the pole dominance is well satisfied. On the other hand, the gluon condensate plays a tiny important role, the operator product expansion is well convergent. It is reliable to extract the masses and pole residues,   which are shown  in Table \ref{Borel-Pole} and Figs.\ref{mass-Borel}-\ref{residue-Borel}.

 The predicted mass $M_P=6.274\pm0.054\,\rm{GeV}$ is in very good agreement with the experimental data $6274.47\pm 0.27\pm 0.17\,\rm{MeV}$ from the Particle Data Group \cite{PDG}, while the predicted mass $M_S=6.702\pm0.060\,\rm{GeV}$ is consistent with other theoretical calculations \cite{GI-1,GI-2,RPMD-EFG,RPMD-GJ,RPMD-ZVR,NRPMD-SR-GKLT-1,NRPMD-SR-GKLT-2,NRPMD-EQ,NRPMD-Monteiro,
 NRPMD-Fulc,NRPMD-Soni,NRPMD-Ortega,NRPMD-LuQF,Latt0909-Gregory-PRL,Latt-Davies,Latt-Mathur-mass,
 BSE-Vary,BSE-WangGL-1,BSE-WangGL-2}.

Combined with our previous work \cite{QCDSR-WZG-BcV}, we can observe that there exist the relations $\sqrt{s^0_V}-M_{V}\approx \sqrt{s^0_P}-M_{P} \approx 0.4\,\rm{GeV}$ and $\sqrt{s^0_A}-M_{A}\approx \sqrt{s^0_S}-M_{S} \approx 0.6\,\rm{GeV}$. Usually, we expect that the energy gaps between the ground states and first radial excitations are about $0.6\,\rm{GeV}$. In practical calculations, we can set the continuum threshold parameter $\sqrt{s_0}$ to be
 any values between the ground state and first radial excitation, i.e. $M_{\rm 1S}+\frac{\Gamma_{\rm 1S}}{2}< \sqrt{s_0}< M_{\rm 2S}-\frac{\Gamma_{\rm 2S}}{2}$, if good QCD sum rules can be obtained, where the 1S and $2S$ stand for the ground state and first radial excitation, respectively. The energy gaps $0.4\,\rm{GeV}$ and $0.6\,\rm{GeV}$ are all make sense.

\begin{table}
\begin{center}
\begin{tabular}{|c|c|c|c|c|c|c|c|}\hline\hline
               &$T^2 (\rm{GeV}^2)$ &$s_0 (\rm{GeV}^2)$ &pole        &$M(\rm{GeV})$    &$f(\rm{GeV})$   \\ \hline

$B_{c}({0}^-)$ &$3.0-4.0$          &$44\pm1$           &$(68-89)\%$ &$6.274\pm0.054$  &$0.371\pm0.037$  \\ \hline

$B_{c}({0}^+)$ &$5.4-6.4$          &$54\pm1$           &$(69-83)\%$ &$6.702\pm0.060$  &$0.236\pm0.017$  \\ \hline

$\widehat{B}_{c}({0}^-)$ &$2.4-3.4$          &$44\pm1$           &$(75-94)\%$ &$6.275\pm0.045$  &$0.208\pm0.015$  \\ \hline

$\widehat{B}_{c}({0}^+)$ &$3.5-4.5$          &$54\pm1$           &$(85-96)\%$ &$6.704\pm0.055$  &$0.119\pm0.006$  \\ \hline\hline
 \end{tabular}
\end{center}
\caption{ The Borel windows, continuum threshold parameters, pole contributions, masses and decay constants  of  the pseudoscalar and scalar  $B_c$ mesons, where the \,$\widehat{}$\, denotes that the radiative $\mathcal{O}(\alpha_s)$ corrections have been neglected.   }\label{Borel-Pole}
\end{table}

As for  the decay constants, even for the pseudoscalar $B_c$ meson, the theoretical values vary in a large range, for example, the values from the full QCD sum rules (QCDSR) \cite{NRPMD-SR-GKLT-2,QCDSR-Chabab1994,QCDSR-Colangelo1993,QCDSR-Aliev-Bc,
QCDSR-Narison-Bc,QCDSR-Schilcher-Bc}, the relativistic quark model (RQM) \cite{GI-2,RPMD-EFG},  the non-relativistic quark model (NRQM)  \cite{NRPMD-EQ,NRPMD-Monteiro},
 the light-front quark model (LFQM) \cite{Decay-LFQM-Tang},
   the lattice non-relativistic QCD (LNQCD) \cite{Jones-latt},
    the  shifted
$N$-expansion method (SNEM) \cite{Ikhdair2006},
      the field correlator method (FCM) \cite{BBS},
      the Bethe-Salpeter equation (BSE) \cite{BSE-WangGL-1}, etc; and we present those values  in Table \ref{Residues-Ref} for clearness. At the present time, it is difficult to say which value is  superior to others.

 The present prediction  $f_P=371\pm37\,\rm{MeV}$ is in very good agreement with the value $371\pm17\,\rm{MeV}$ from the full QCD sum rules \cite{QCDSR-Narison-Bc}. In our previous work, we obtain the values $f_V=384\pm32\,\rm{MeV}$
and $f_A=373\pm25\,\rm{MeV}$ for the vector and axial-vector $B_c$ mesons, respectively \cite{QCDSR-WZG-BcV}. Our calculations indicate that $f_P\approx f_V \approx f_A > f_S$.
While in the QCD sum rule combined with the heavy quark effective theory up to the order $\alpha_s^3$,
the decay constants have the relations $\tilde{f}_P=f_P> f_V>f_S> \tilde{f}_S>f_A$ \cite{Decay-SR-HQEFT-XiaoZJ}, where the decay constants  $\tilde{f}_P$ and $ \tilde{f}_S$ are defined by
\begin{eqnarray}\label{decay-const-defin-2}
 \langle 0|J_P(0)|P(p)\rangle&=&\tilde{f}_PM_P\, , \nonumber \\
 \langle 0|J_S(0)|S(p)\rangle&=&\tilde{f}_SM_S\, .
 \end{eqnarray}
From Eq.\eqref{decay-const-defin-1} and Eq.\eqref{decay-const-defin-2}, we can obtain the relations,
\begin{eqnarray}
\tilde{f}_P&=&f_P\frac{M_P}{m_b+m_c}\, , \nonumber\\
\tilde{f}_S&=&f_S\frac{M_S}{m_b-m_c}\, ,
\end{eqnarray}
it is obvious that $\tilde{f}_P>f_P$ and $\tilde{f}_S<f_S$, which are in contrary to the relations obtained in Ref.\cite{Decay-SR-HQEFT-XiaoZJ}, so no definite conclusion can be obtained. Naively, we expect that the vector mesons have larger decay constants than the corresponding pseudoscalar mesons \cite{WZG-decay-EPJC}.

If we neglect the radiative $\mathcal{O}(\alpha_s)$ corrections (in other words, the next-to-leading order contributions), the same input parameters would lead to too large hadron masses, we have to choose the energy scales $\mu=2.1\,\rm{GeV}$ and $2.2\,\rm{GeV}$ for the pseudoscalar and scalar $B_c$ mesons, respectively. Then we refit the Borel parameters, the corresponding pole contributions,  masses and decay constants are  given explicitly  in Table \ref{Borel-Pole}. Form the Table, we can see explicitly that the predicted masses change slightly, while the predicted decay constants change greatly, the decay constants without the radiative $\mathcal{O}(\alpha_s)$ corrections only count for about $56\%$ of the corresponding ones with the radiative $\mathcal{O}(\alpha_s)$  corrections. The radiative $\mathcal{O}(\alpha_s)$ corrections play an important role, we should take it into account.

\begin{table}
\begin{center}
\begin{tabular}{|c|c|c|c|c|c|c|c|}\hline\hline
      &$f_P(\rm{MeV})$     &References       \\ \hline

QCDSR &$460\pm60$          &\cite{NRPMD-SR-GKLT-2}   \\ \hline

QCDSR &$300\pm65$          &\cite{QCDSR-Chabab1994}   \\ \hline

QCDSR &$360\pm60$          &\cite{QCDSR-Colangelo1993}  \\ \hline

QCDSR &$270\pm30$          &\cite{QCDSR-Aliev-Bc}   \\ \hline

QCDSR &$371\pm17$          &\cite{QCDSR-Narison-Bc}   \\ \hline

QCDSR &$528\pm 19$         &\cite{QCDSR-Schilcher-Bc}  \\ \hline

QCDSR &$371\pm37$          &This work   \\ \hline

RQM   &$410\pm40$          &\cite{GI-2}  \\ \hline

RQM   &$433$               &\cite{RPMD-EFG}  \\ \hline

NRQM  &$498$               &\cite{NRPMD-EQ}  \\ \hline

NRQM  &$440$               &\cite{NRPMD-Monteiro}  \\ \hline

LFQM  &$523\pm 62$         &\cite{Decay-LFQM-Tang}  \\ \hline

LNQCD &$420\pm 13$         &\cite{Jones-latt}  \\ \hline

LNEM  &$315^{+26}_{-50}$   &\cite{Ikhdair2006}  \\ \hline

FCM   &$438\pm 10$         &\cite{BBS} \\ \hline

BSE   &$322 \pm 42$        &\cite{BSE-WangGL-1} \\ \hline\hline

 \end{tabular}
\end{center}
\caption{ The decay constant  of  the pseudoscalar $B_c$ meson from different theoretical works.   }\label{Residues-Ref}
\end{table}

 The pure leptonic decay widths $\Gamma_{\ell \bar{\nu}_{\ell}}$ of the pseudoscalar and scalar  $B_c$ mesons can be written as,
\begin{eqnarray}
\Gamma_{\ell \bar{\nu}_{\ell}} &=& \frac{G_F^2}{8\pi}|V_{bc}|^2f_{P/S}^2M_{P/S}M_{\ell}^2 \left(1-\frac{M_{\ell}^2}{M_{P/S}^2} \right)^2 \, ,
\end{eqnarray}
where the leptons $\ell=e,\mu,\tau$, the Fermi constant $G_F=1.16637\times10^{-5} \,\rm{GeV}^{-2}$, the CKM matrix element  $V_{cb}=40.8\times 10^{-3}$, the masses of the leptons  $m_e=0.511\times 10^{-3}\,\rm{GeV}$, $m_\mu=105.658\times 10^{-3}\,\rm{GeV}$,
$m_\tau=1776.86\times 10^{-3}\,\rm{GeV}$, the life time of the $B_c$ meson $\tau_{B_c}=0.510\times 10^{-12}\,s$ from the Particle Data Group  \cite{PDG}. We take  the masses and decay constants of the pseudoscalar and scalar  $B_c$ mesons  from the QCD sum rules to obtain the partial  decay widths,
\begin{eqnarray}
\Gamma_{P\to e \bar{\nu}_{e}}&=&2.03 \times 10^{-12}\,\rm{eV}\, ,  \nonumber \\
\Gamma_{P\to \mu \bar{\nu}_{\mu}}&=&8.68  \times 10^{-8}\,\rm{eV}\, ,  \nonumber \\
\Gamma_{P\to \tau \bar{\nu}_{\tau}}&=&2.08 \times 10^{-5}\,\rm{eV}\, , \nonumber \\
\Gamma_{S\to e \bar{\nu}_{e}}&=&8.78 \times 10^{-13}\,\rm{eV}\, ,  \nonumber \\
\Gamma_{S\to \mu \bar{\nu}_{\mu}}&=&3.75 \, \times 10^{-8}\,\rm{eV}\, , \nonumber \\
\Gamma_{S\to \tau \bar{\nu}_{\tau}}&=&9.18 \,\times 10^{-6}\,\rm{eV} \, ,
\end{eqnarray}
and the branching fractions,
\begin{eqnarray}
{\rm Br}_{P\to e \bar{\nu}_{e}}&=&1.57 \times 10^{-9} \, ,  \nonumber \\
{\rm Br}_{P\to \mu \bar{\nu}_{\mu}}&=&6.73  \times 10^{-5}\, ,  \nonumber \\
{\rm Br}_{P\to \tau \bar{\nu}_{\tau}}&=&1.61 \times 10^{-2}\,  .
\end{eqnarray}
The largest branching fractions of the $B_c(0^-)\to  \ell \bar{\nu}_\ell$  are of the order $10^{-2}$, the tiny branching fractions  maybe  escape experimental  detections. By precisely measuring the branching fractions, we can examine the theoretical calculations strictly, although it is a hard work.

\section{Conclusion}
In this work, we extend our previous works on the vector and axialvector $B_c$ mesons to  investigate  the pseudoscalar and scalar $B_c$ mesons with the full QCD sum rules by including next-to-leading order corrections and choose the  updated input parameters. In calculating the next-to-leading order corrections, we  use optical theorem (or Cutkosky's rule) to obtain the QCD spectral densities straightforwardly, and resort to the dimensional regularization to regularize both the ultraviolet and infrared divergences, which  are canceled out with each other separately, the total  QCD spectral densities have neither  ultraviolet divergences nor infrared  divergences. Then we calculate  the gluon condensate contributions and reach the QCD sum rules.   We take the experimental data as guides to choose the suitable Borel parameters and continuum threshold parameters,   and make reasonable predictions for the masses and decay constants, therefore the pure leptonic decay widths, which can  be confronted to the experimental data in the future to examine the theoretical calculations or extract the decay constants, which are fundamental input parameters in the high energy physics.

\section*{Acknowledgements}
This work is supported by National Natural Science Foundation, Grant Number 12175068.

\section*{Appendix}
At first, we write down all the elementary integrals involving  the vertex corrections,
\begin{eqnarray}
V_{ab}(s)&=& 16\pi^2\int_0^1 dx \int_0^{1-x}dy \int \frac{d^D k_E}{(2\pi)^D} \frac{x^a y^b \Gamma(3)}{\left[k^2_E+(x p_1+y p_2)^2\right]^3}  \, ,\nonumber\\
V(s)&=& 16\pi^2\left(1-\frac{1}{2}\varepsilon_{\rm UV}\right)\int_0^1 dx \int_0^{1-x}dy \int \frac{d^D k_E}{(2\pi)^D}
\frac{k^2_E \Gamma(3)}{\left[k^2_E+(x p_1+y p_2)^2\right]^3} \, ,
\end{eqnarray}
and accomplish all the integrals to acquire  the analytical expressions,
\begin{eqnarray}
V_{00}(s)&=& \frac{1}{\sqrt{\lambda(s,m_b^2,m_c^2)}}\left\{ -\log\left(\frac{1+\omega}{1-\omega}\right)\left(\frac{1}{\varepsilon_{\rm IR}}+\log\frac{s}{4\pi\mu^2}+\gamma \right) + \frac{\log^2(1-\omega_1^2)}{4}-\log^2(1+\omega_1) \right. \nonumber \\
&&+ \frac{\log^2(1-\omega_2^2)}{4}-\log^2(1+\omega_2) +2\log(\omega_1+\omega_2)\log\left(\frac{1+\omega}{1-\omega}\right)-\log\omega_1\log\left(\frac{1+\omega_2}{1-\omega_2}\right) \nonumber \\
&&\left.-\log\omega_2\log\left(\frac{1+\omega_1}{1-\omega_1}\right)-{\rm Li_2}\left( \frac{2\omega_1}{1+\omega_1} \right) -{\rm Li_2}\left( \frac{2\omega_2}{1+\omega_2} \right)+\pi^2\right\} \, ,\nonumber \\
&=&\overline{V}_{00}(s)- \frac{1}{\sqrt{\lambda(s,m_b^2,m_c^2)}}\log\left(\frac{1+\omega}{1-\omega}\right)\left(\frac{1}{\varepsilon_{\rm IR}}+\log\frac{s}{4\pi\mu^2}+\gamma \right) \, , \nonumber
\end{eqnarray}
\begin{eqnarray}
V_{10}(s)&=& \frac{1}{s}\left\{\frac{1}{2}\log\left(\frac{1-\omega_1^2}{1-\omega_2^2}\right)-\frac{1}{\omega_2}\log\left(\frac{1+\omega}{1-\omega} \right)+\log\frac{\omega_2}{\omega_1}\right\} \, ,\nonumber \\
V_{01}(s)&=& V_{10}(s)|_{\omega_1 \leftrightarrow \omega_2} \, ,\nonumber \\
V_{20}(s)&=& \frac{1}{2s}\left\{-\frac{\omega_1\omega_2}{\omega_1+\omega_2}\log\left(\frac{1+\omega}{1-\omega} \right)-\frac{\omega_1}{\omega_2(\omega_1+\omega_2)}\log\left(\frac{1+\omega}{1-\omega} \right)+   \frac{\omega_1}{\omega_1+\omega_2}\log\left(\frac{1-\omega_1^2}{1-\omega_2^2}\right)\right.\nonumber\\
&&\left.+\frac{2\omega_1}{\omega_1+\omega_2}\log\frac{\omega_2}{\omega_1}+1\right\} \, ,\nonumber \\
V_{02}(s)&=&V_{20}(s)|_{\omega_1 \leftrightarrow \omega_2} \, ,\nonumber \\
V_{11}(s)&=& \frac{1}{2s}\left\{\frac{\omega_1\omega_2}{\omega_1+\omega_2}\log\left(\frac{1+\omega}{1-\omega} \right)-  \frac{\omega_1-\omega_2}{2(\omega_1+\omega_2)}\log\left(\frac{1-\omega_1^2}{1-\omega_2^2}\right)-\frac{1}{\omega_1+\omega_2}\log\left(\frac{1+\omega}{1-\omega} \right)\right.\nonumber\\
&&\left.+\frac{\omega_1}{\omega_1+\omega_2}\log\frac{\omega_1}{\omega_2}+\frac{\omega_2}{\omega_1+\omega_2}\log\frac{\omega_2}{\omega_1}-1\right\} \, ,\nonumber
\end{eqnarray}
\begin{eqnarray}
V(s)&=& \frac{1}{\varepsilon_{\rm UV}}+\log\frac{4\pi\mu^2}{s}-\gamma+2-\frac{2\omega_1\omega_2}{\omega_1+\omega_2}\log\left(\frac{1+\omega}{1-\omega} \right)
-\frac{\omega_2}{\omega_1+\omega_2}\log(1-\omega_1^2)\nonumber \\
&&-\frac{\omega_1}{\omega_1+\omega_2}\log(1-\omega_2^2)-2\frac{\omega_1\log\omega_1+\omega_2\log\omega_2}{\omega_1+\omega_2}+2\log(\omega_1+\omega_2)\, , \nonumber\\
&=& \overline{V}(s)+ \frac{1}{\varepsilon_{\rm UV}}+\log\frac{4\pi\mu^2}{s}-\gamma+2\, ,
\end{eqnarray}
where
\begin{eqnarray}
\omega_1&=&\frac{\sqrt{\lambda(s,m_b^2,m_c^2)}}{s+m_b^2-m_c^2} \, ,\nonumber\\
\omega_2&=&\frac{\sqrt{\lambda(s,m_b^2,m_c^2)}}{s+m_c^2-m_b^2} \, ,\nonumber\\
M&=&\frac{m_b+m_c}{m_b-m_c}\, ,\nonumber\\
{\rm Li_2}(x)&=&-\int_0^x dt \frac{\log(1-t)}{t} \, .
\end{eqnarray}

Then we introduce  the notation
\begin{eqnarray}
\int d ps&=& \int \frac{d^{D-1}\vec{k}}{2E_k}\frac{d^{D-1}\vec{p}_1}{2E_{p_1}}\frac{d^{D-1}\vec{p}_2}{2E_{p_2}}\delta^D(p-k-p_1-p_2) \, ,\nonumber
\end{eqnarray}
for simplicity, and write down the elementary three-body phase-space integrals,
\begin{eqnarray}
 R_{11}(s)&=& \frac{s m_b^2}{\pi^2\sqrt{\lambda(s,m_b^2,m_c^2)}}(2\pi)^{-4\varepsilon_{\rm IR}}\mu^{-2\varepsilon_{\rm IR}}\int d ps \frac{1}{(k \cdot p_1 )^2}\nonumber\\
&=& \frac{1}{2\varepsilon_{\rm IR}}-\log4\pi+\gamma-1+\log\frac{\sqrt{\lambda(s,m_b^2,m_c^2)}^3}{m_bm_cs\mu^2}
-\frac{s+m_b^2-m_c^2}{2\sqrt{\lambda(s,m_b^2,m_c^2)}} \log\left(\frac{1+\omega_1}{1-\omega_1}\right) \nonumber \\
&&-\frac{m_b^2-m_c^2}{\sqrt{\lambda(s,m_b^2,m_c^2)}} \log\left(\frac{1+\omega_1}{1-\omega_1}\right) -\frac{s-m_b^2+m_c^2}{\sqrt{\lambda(s,m_b^2,m_c^2)}} \log\left(\frac{1+\omega}{1-\omega}\right) \nonumber \\
&=&\overline{ R}_{11}(s)+\frac{1}{2\varepsilon_{\rm IR}}-\log4\pi+\gamma-1+\log\frac{\sqrt{\lambda(s,m_b^2,m_c^2)}^3}{m_bm_cs\mu^2} \, , \nonumber \\
R_{22}(s)&=&R_{11}(s)|_{m_b\leftrightarrow m_c} \, , \nonumber
\end{eqnarray}
\begin{eqnarray}
R_{12}(s)&=& \frac{s  }{\pi^2\sqrt{\lambda(s,m_b^2,m_c^2)}}(2\pi)^{-4\varepsilon_{\rm IR}}\mu^{-2\varepsilon_{\rm IR}}\int d ps \frac{1}{k\cdot p_1 k\cdot p_2 }\nonumber\\
&=&\frac{1}{\sqrt{\lambda(s,m_b^2,m_c^2)}} \left\{\log\left(\frac{1+\omega}{1-\omega}\right)\left[ \frac{1}{\varepsilon_{\rm IR}}-2\log4\pi+2\gamma-2+2\log\frac{\sqrt{\lambda(s,m_b^2,m_c^2)}^3}{m_bm_cs\mu^2}\right] \right. \nonumber \\
&&-2\log\frac{m_b}{m_c}\log\left(\frac{M+\omega}{M-\omega}\right)-\log^2\left(\frac{1+\omega}{1-\omega}\right)+2\log\frac{s}{\bar{s}}\log\left(\frac{1+\omega}{1-\omega}\right)
-4{\rm Li_2}\left( \frac{2\omega}{1+\omega}\right) \nonumber\\
&&+2{\rm Li_2}\left( \frac{\omega-1}{\omega-M}\right) +2{\rm Li_2}\left( \frac{\omega-1}{\omega+M}\right)-2{\rm Li_2}\left( \frac{\omega+1}{\omega-M}\right)-2{\rm Li_2}\left( \frac{\omega+1}{\omega+M}\right)-\frac{1}{2}{\rm Li_2}\left( \frac{1+\omega_1}{2}\right)\nonumber\\
&&\left.-\frac{1}{2}{\rm Li_2}\left( \frac{1+\omega_2}{2}\right)-{\rm Li_2}\left( \omega_1\right)-{\rm Li_2}\left( \omega_2\right)+\frac{\log2\log\left[(1+\omega_1)(1+\omega_2) \right]}{2}-\frac{\log^2 2}{2}+\frac{\pi^2}{12}\right\}\, ,\nonumber\\
&=&\overline{R}_{12}(s)+\frac{1}{\sqrt{\lambda(s,m_b^2,m_c^2)}} \log\left(\frac{1+\omega}{1-\omega}\right)\left[ \frac{1}{\varepsilon_{\rm IR}}-2\log4\pi+2\gamma-2+2\log\frac{\sqrt{\lambda(s,m_b^2,m_c^2)}^3}{m_bm_cs\mu^2}\right] \, , \nonumber
\end{eqnarray}
\begin{eqnarray}
R^1_{12}(s)&=& \frac{s  }{\pi^2\sqrt{\lambda(s,m_b^2,m_c^2)}}\int d ps \frac{s-K^2}{k\cdot p_1 k\cdot p_2 }\nonumber\\
&=&\frac{s}{\sqrt{\lambda(s,m_b^2,m_c^2)}} \left\{\log^2(1-\omega)-\log^2(1+\omega) +2\log\frac{2s}{\bar{s}}\log\left(\frac{1+\omega}{1-\omega}\right)
+2{\rm Li_2}\left(\frac{1-\omega}{2}\right)\right. \nonumber \\
&&\left.-2{\rm Li_2}\left(\frac{1+\omega}{2}\right)+2{\rm Li_2}\left(\frac{1+\omega}{1+M}\right)+2{\rm Li_2}\left(\frac{1+\omega}{1-M}\right)
-2{\rm Li_2}\left(\frac{1-\omega}{1-M}\right)-2{\rm Li_2}\left(\frac{1-\omega}{1+M}\right)\right\} \, ,\nonumber \\
R^2_{12}(s)&=& \frac{s}{\pi^2\sqrt{\lambda(s,m_b^2,m_c^2)}}\int d ps \frac{\left(s-K^2\right)^2}{k\cdot p_1 k\cdot p_2 }\nonumber\\
&=&\frac{s^2}{\sqrt{\lambda(s,m_b^2,m_c^2)}} \left\{\log^2(1-\omega)-\log^2(1+\omega) +2\log\frac{4s}{\bar{s}}\log\left(\frac{1+\omega}{1-\omega}\right)
+2{\rm Li_2}\left(\frac{1-\omega}{2}\right)\right. \nonumber \\
&&-2{\rm Li_2}\left(\frac{1+\omega}{2}\right)+2{\rm Li_2}\left(\frac{1+\omega}{1+M}\right)+2{\rm Li_2}\left(\frac{1+\omega}{1-M}\right)
-2{\rm Li_2}\left(\frac{1-\omega}{1-M}\right)-2{\rm Li_2}\left(\frac{1-\omega}{1+M}\right) \, \nonumber \\
&&\left. +\frac{2\omega \bar{s}}{s} -\frac{\bar{s}}{s}(1+\omega^2)\log\left(\frac{1+\omega}{1-\omega}\right) \right\} \, ,
\end{eqnarray}
where $\bar{s}=s-(m_b-m_c)^2$.

\end{document}